\documentclass[11pt,a4paper]{article}
\usepackage{jcappub}
\usepackage{mathrsfs}
\usepackage{epsfig}
\usepackage{cite}
\usepackage{amssymb}
\usepackage{rotating}

\newcommand{\D}{\displaystyle}

\def\aa{{\it Astron. Astrophys.} \,}

\def\apj{{\it ApJ \,}}
\def\apjs{{\it Ap. J. Supp.} \,}

\def\prd{{\it Phys. Rev. D.} \,}
\def\mn{{\it MNRAS} \,}

\def\pasj{{\it Pub. Astron. Soc. Japan} \,}

\title{CMB Anisotropy Due to Filamentary Gas: Power Spectrum and Cosmological 
Parameter Bias}

\author[1]{Meir Shimon,}\emailAdd{meirs@wise.tau.ac.il}
\author[1]{Sharon Sadeh,}
\author[1,2]{Yoel Rephaeli}
\affiliation[1]{School of Physics and Astronomy, Tel Aviv University, Tel Aviv 
69978, Israel}
\affiliation[2]{Center for Astrophysics and Space Sciences, University of California, 
San Diego, La Jolla, CA, 92093}

\abstract{Hot gas in filamentary structures induces CMB aniostropy through 
the SZ effect. Guided by results from N-body simulations, we model the 
morphology and gas properties of filamentary gas and determine the power 
spectrum of the anisotropy. Our treatment suggests that power levels 
can be an appreciable fraction of the cluster contribution at multipoles 
$\ell\lesssim 1500$. Its spatially irregular morphology and larger 
characteristic angular scales can help to distinguish this SZ signature 
from that of clusters. In addition to intrinsic interest in this most 
extended SZ signal as a probe of filaments, its impact on cosmological 
parameter estimation should also be assessed. We find that filament `noise' 
can potentially bias determination of $A_s$, $n_s$, and $w$ (the 
normalization of the primordial power spectrum, the scalar index, 
and the dark energy equation of state parameter, respectively) by more than 
the nominal statistical uncertainty in Planck SZ survey data. 
More generally, when inferred from future optimal cosmic-variance-limited 
CMB experiments, we find that virtually all parameters will be biased by 
more than the nominal statistical uncertainty estimated for these next 
generation CMB experiments.} 


\begin{document}

\maketitle

\flushbottom

\section{Introduction}

As much as $30-40\%$ of the baryon mass density of the universe is 
thought to be in a phase of warm gas distributed along large-scale 
filamentary structures connecting clusters and superclusters of 
galaxies. These structures are commonly identified in numerical 
hydrodynamic simulations [1,2].
Estimated temperature and density ranges are $10^5<T<10^7\,K$ and 
$4\cdot 10^{-6}<n<10^{-4}\,cm^{-3}$, respectively [1,3]. Emission from 
this gas would be expected in the soft X-ray region, yet its detection 
constitutes a major observational challenge due to the very low surface 
brightness and confusion by other extended sources such as clusters and 
groups. 

Observational evidence for this so-called warm-hot intergalactic 
medium (WHIM) is still quite scant, mostly from measurements of 
absorption lines in the spectra of background AGN. These include e.g.
measurement of an O VIII absorption system in the spectrum of a quasar 
observed in the Virgo cluster region, possibly due to WHIM around the 
cluster [4], detection of WHIM associated with the Coma cluster by 
identifying Ne IX and O VIII absorption lines in the spectrum of X 
Comae, an active galactic nucleus (AGN) located behind the Coma 
cluster [5], observation of X-ray emission from the WHIM in the 
Shapley supercluster region [6], in a high Galactic latitude ROSAT 
field [7], and in the outskirts of the Coma cluster [8], as 
measured by X-ray Multi-Mirror Mission (XMM-Newton).

Morphological and statistical studies of filamentary dark matter 
(hereafter DM) structures in N-body simulations were conducted 
by e.g. [9,10], who addressed such properties as filament lengths, 
masses, and shapes, the DM density distribution along and across 
filaments, and the number of filaments per cluster. More recently, 
Klar \& M\"{u}cket [11] have performed hydrodynamical simulations, 
studying in detail hydro- and thermodynamical properties of 
intra-filament (IF) gas, including its temporal evolution.

The WHIM may obviously have other detectable signatures, most 
important of which would be its impact on the spatial structure of the 
cosmic microwave background (CMB): 
Compton scattering of this radiation field by free electrons 
induces an additional component to the anisotropy due to the SZ 
effect in virialized systems, mostly clusters of galaxies. The 
WHIM-induced contribution could be appreciable even though the 
gas is cooler and significantly thinner than intracluster (IC) gas, 
owing to the long path lengths associated with the filamentary WHIM. 
CMB anisotropy induced by IC gas - the dominant source of (secondary) 
anisotropy on angular scales of a few arcminutes - has been 
extensively studied, e.g. [12-14], resulting in detailed predictions 
of its power spectrum and cluster number counts. 

In contrast, the related anisotropy due IF gas has attracted little 
attention, likely due to its presumed low level, and the scant 
observational information on properties of IF gas. These were 
explored by cosmological hydrodynamical simulations, first by 
Cen \& Ostriker [1]. An analytical study of the power spectrum 
of the Sunyaev-Zeldovich (SZ) effect in intergalactic gas (IGG) 
was carried out in [15], who assumed a log-normal random baryon 
field for the IGG. In this scenario the contribution to SZ power 
levels is likely to originate in both larger and smaller structures, 
such as filaments, halos, and possibly even smaller objects. 
More recently, in their analysis of the Santa Fe Light Cone 
Simulation, Hallman et al. [2] identified groups and clusters down 
to $5\times10^{13}$ solar masses. Subtracting the SZ signal from 
these systems, they mapped the residual magnitude of the 
Comptonization parameter $y$, finding rather high levels, 
typically a third of the mean level expected in clusters. 

In this paper we describe the results of an analytical calculation of
the SZ power signature of IF gas on the CMB and its impact on cosmological 
parameter inference from the CMB. The calculation of the power spectrum 
consists of a simple morphological modeling of filaments and their 
cosmic abundance, an approach which is a direct extension of previous 
work on SZ cluster-induced anisotropy. Given the current limited 
knowledge of IGG properties and the morphological complexity of gaseous 
IG structures, we are inevitably compelled to model them in the simplest 
manner possible, which nonetheless provides us with the flexibility of 
testing and exploring several inherent limitations associated with 
cosmological simulations. The calculation of the bias that the 
filaments-induced CMB anisotropy introduces on cosmological parameter 
inference is standard and follows the conventional Fisher information 
matrix methodology. 

The structure of this paper is as follows: 
Following the Introduction, Sec. 1, we detail in Sec. 2 the simple 
model we adopted for the morphological and physical properties of the 
IF gas. We further describe the cosmological abundance of filaments, 
and lay out the computation methodology of the power spectrum. In Sec. 3 
we describe the calculation of the cosmological parameters bias. 
Results of our work are presented in Sec. 4, discussed in Sec. 5, and 
summarized in Sec. 6.

\section{Power Spectrum Calculation}

\subsection{Filament Model and Comptonization Parameter}

The complex web of filamentary structure is difficult to model, but for 
our purposes here we can roughly describe filaments as isolated entities 
whose main gaseous sections have cylindrical morphology. Using cylindrical 
coordinates in the frame of the filament, and assuming a uniform IF gas 
temperature and density, the Comptonization parameter at position 
$\vec{r}=(r,\phi,z)$ can be expressed in terms of the cross sectional 
(2D) radius $R$ of the filament, its length along the symmetry axis, $L$, 
and Heaviside step functions  
\begin{equation}
y(r,\phi,z)=y_0\Theta(R-|r|)\Theta(L/2-|z|),
\end{equation} 
where $y_0\equiv\int\D\frac{k_B T_e}{m_e c^2}\,\sigma_T\,n_e\,d\ell$, with 
$T_e$, $n_e$, and $\sigma_T$ denoting the electron temperature, density and 
the Thomson cross section, respectively, and $d\ell$ a line element along 
the line of sight (los); 
the remaining quantities have their usual meanings. In our numerical 
calculations we take $T_e =1\,keV$ and $n_e = 4.3\times 10^{-6}\,cm^{-3}$, 
but these are just fiducial values gauging their allowed ranges. As in 
virialized systems, the gas mass fraction is taken to be a constant fraction 
$f_g$ of the total mass, $M$; denoting the hydrogen mass fraction in the gas 
as $X$, the (total) IF  mass density is 
\begin{equation}
\rho_m=\frac{n_e m_p (X+1)}{2 f_g},
\end{equation}
where $m_p$ is the proton mass, and $X=0.59$ (in fully ionized gas). We 
take $f_g = 0.12$, a typical value for the cosmological baryon fraction 
of the mass density, which also proves to be consistent with the assumption 
that a fraction $30-40\%$ of the total baryonic matter in the universe is 
in filaments (as is discussed later in this section). The inferred filament 
length 
\begin{equation}
L=\frac{M}{\pi R^2\rho_m}
\label{eq:filh}
\end{equation}
is expressed in terms of $R$, which is scaled to a value of 
$1.5\,Mpc\cdot h^{-1}$, typical over a wide range of filament masses [9]. 

\subsection{Temperature Anisotropy}

We first compute the Comptonization parameter Fourier transform, required
for estimating the power spectrum of the temperature anisotropy.
With the relations detailed in the previous subsection, the Fourier 
transformed $y$ in the filament frame can be calculated analytically, writing 
\begin{eqnarray}
& &\vec{r}=r\cos{\phi}\vec{\hat{x}}+r\sin{\phi}\vec{\hat{y}}+z\vec{\hat{z}}\equiv 
r\vec{\hat{n}}+z\vec{\hat{z}} \nonumber \\
& & \vec{k}=k\cos{\phi}'\vec{\hat{x}}+k\sin{\phi}'\vec{\hat{y}}
+k_z\vec{\hat{z}}\equiv k\vec{\hat{n}'}+k_z\vec{\hat{z}}, 
\end{eqnarray}
so that $e^{i\vec{k}\cdot\vec{r}}=e^{ikr\cos{\phi}}e^{ik_zz}$, 
where $r=\sqrt{x^2+y^2}$ and $k=\sqrt{k_x^2+k_y^2}$.
Consequently, 
\begin{eqnarray}
& &\tilde{y}(\vec{k})=\int_0^{2\pi}\int_{-\infty}^{\infty}\int_0^{\infty} 
y_0\Theta(R-|r|)\Theta(L/2-|z|)e^{ikr\cos{\phi}}e^{ik_zz}\,r\,dr\,dz\,
d\phi \nonumber \\
& &=\frac{4\pi y_0 R}{k k_z} J_1(kR) \sin{\left(\frac{h\,k_z}{2}\right)}.
\label{eq:ft}
\end{eqnarray}

The desired expression is the Fourier transform of $y$ in the observer 
frame, which is chosen such that the axis of symmetry of the filament 
forms an angle $\theta$ with the los to the filament, as illustrated in 
Figure 1. Transforming the components of the wave vector $\vec{k}$ in the 
filament frame into the observer frame $\vec{k'}$ is easily done as  

\begin{eqnarray}
& & k_x=k'_x\cos{\theta}-k'_z\sin{\theta}\nonumber\\
& & k_y=k'_y\\
& & k_z=k'_x\sin{\theta}+k'_z\cos{\theta}\nonumber, 
\end{eqnarray}
where primed and unprimed quantities correspond to the observer and 
filament frames, respectively. Substituting for the unprimed coordinates 
in equation~(\ref{eq:ft}) yields the Fourier transformed profile of the 
filament y-parameter in terms of the components of the wave vector in the 
observer frame,
\begin{equation}
\tilde{y}(\vec{k'})=\frac{4\pi y_0 RJ_1\left[R\sqrt{(k_x'\cos{\theta}-
k'_z\sin{\theta})^2+k_y'^2}\right]\sin\left[(k'_x\sin{\theta}+
k'_z\cos{\theta})(L/2)\right]}{\sqrt{(k'_x\cos{\theta}-k'_z\sin{\theta})^2
+k_y'^2}(k'_x\sin{\theta}+k'_z\cos{\theta})}.
\label{eq:ft2}
\end{equation}

\begin{figure}
\centering
\epsfig{file=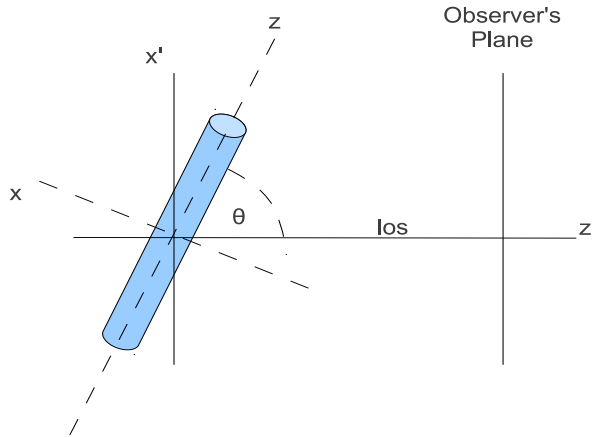, width=8cm, height=6cm,clip=}
\caption{The observer (primed) and filament (unprimed) reference frames.
The filament symmetry axis is inclined by an angle $\theta$ to the los
(figure adapted from [16]).}
\label{fig:geom}
\end{figure}

\subsection{Power Spectrum}

The 3-dimensional power spectrum due to a single filament, as would be 
measured by the observer, can be derived by taking the ensemble average of 
$y^2(k)$. 
The 3- and 2-dimensional power spectra can be related by expressing the latter 
as an angular Fourier integral over the projected (angular) profile of the 
Comptonization parameter, which, in turn, can be evaluated by means of 
a los integral over the 3-dimensional y-parameter. Implementing then 
the flat sky approximation, one arrives at the following expression for 
the 2-dimensional angular power spectrum [17]:
\begin{equation}
P_2(\vec{l})=\frac{P(l/D_A)}{D_A^2},
\end{equation}
where $D_A$ denotes the comoving angular diameter distance, $P(k=l/D_A)$ 
is the 3-dimensional power spectrum evaluated at wavenumber $k=l/D_A$, and 
$\vec{l}\equiv(l_x,l_y)$ is a two dimensional vector in a flat patch of sky. 
Global homogeneity and isotropy imply that the power spectrum depends only 
on the magnitude of the vector $l\equiv\sqrt{l_x^2+l_y^2}$. Explicitly,
\begin{equation}
P(l_x,l_y)=\left[\frac{4\pi y_0 RJ_1\left[R\sqrt{(l_x/D_A)^2\cos^2{\theta}
+(l_y/D_A)^2}\right]\sin\left[l_x\sin{\theta}\,(h/2)/D_A\right]}
{\sqrt{(l_x/D_A)^2\cos^2{\theta}+(l_y/D_A)^2}(l_x/D_A)\sin{\theta}}\right]^2.
\end{equation}
In order to find the power spectrum at a given $l$, the contributions from 
all $(l_x,l_y)$ pairs such that $l_y=\sqrt{l^2-l_x^2}$ must be summed up.
This can be easily carried out by working in polar coordinates, in which 
$(l_x,l_y)=(l\cos{\phi},l\sin{\phi})$. Since the power spectrum is quadratic 
in both components of $\vec{l}$, it suffices to integrate over the first 
quadrant of the Cartesian coordinate system and multiply the result by a 
factor 4. 

The resulting power spectrum has to be convolved with a representative mass 
function in order to calculate  the integrated spectrum of the population of 
filaments.

\subsection{Filament Mass Function}

Shen et al. [18] derived mass functions for sheets, filaments, and haloes using 
an excursion set model with a mass-dependent moving barrier. In this formalism, 
filaments are generated whenever the random walk crosses a first barrier
to form a sheet, followed by a second barrier cross, creating the filament. 
The resulting mass function is described by the fitting formula
\begin{equation}
\nu f(\nu)=\sqrt{\frac{\nu}{2\pi}}e^{-\nu(1+\beta\nu^{-\alpha})^2/2}
\left\lbrace 1+\frac{\beta}{\nu^{\alpha}}\left[1-\alpha+\frac{\alpha(\alpha-1)}{2!}
+\cdots\right]\right\rbrace,
\end{equation}
where $\nu\equiv\left[\delta_{c}(z)/\sigma_M\right]^2$. Here $\delta_c$ and 
$\sigma_M$ denote the usual critical overdensity for spherical collapse and 
the mass variance at mass scale $M$, respectively. For filaments, Shen et al. 
provide the fitting parameters $\alpha=0.28$ and $\beta=-0.012$, for which 
the series within the brackets converges to a value of $0.654$. Consequently, 
the mass function assumes the form
\begin{equation}
\nu f(\nu)=\sqrt{\frac{\nu}{2\pi}}e^{-\nu(1-0.012\nu^{-0.28})^2/2}
\left(1-\frac{7.85\cdot 10^{-3}}{\nu^{0.28}}\right).
\end{equation}

The integrated power spectrum due to a population of filaments can now be 
computed with the usual expression
\begin{equation}
C_{\ell}=g^2(\nu)\int_z\int_M dz\,\frac{dV}{dz}\,dM\,\frac{dn(M,z)}{dM}\,P_2(l),
\label{eq:cl}
\end{equation}
where $g(\nu)$ denotes the spectral distortion of the CMB due to the 
SZ effect, and 
\begin{equation}
\frac{dn(M,z)}{dM}\equiv f(\nu)\frac{\bar{\rho}}{M}\frac{d\nu}{dM},
\label{eq:filmf}
\end{equation}
where $\bar{\rho}$ denotes the mean background matter density. 
We integrated the mass function over the redshift and mass ranges of 
$0.01\le z\le 6$ and 
$10^{12}\,M_{\odot}\cdot h^{-1}\le M\le 10^{16}\,M_{\odot}\cdot h^{-1}$, 
respectively. Note that the total baryonic mass contained in the filament 
population described by the mass function should 
constitute a fraction $30-40\%$ of the total baryonic mass. Integrating 
the expression $\int n(M,z) M dM$ over the relevant mass range and dividing
it by a volume element occupied in a redshift shell $\Delta z$ at $z\approx 0$ 
provides the total filament mass density in the present universe, whereas its 
baryonic density can be assessed by multiplying this quantity by the presumed 
gas fraction $f_g=0.12$. The total cosmic baryonic density in the present 
universe can be evaluated as the product $\Omega_b\rho_{c_0}$. We verified that
the baryonic density contributed by the filament population constitutes indeed 
a fraction $30-40\%$ of that implied by $\Omega_b$, or specifically $\sim 32\%$ 
for $f_g=0.12$ and the choice of cosmological parameters detailed in 
Section ~\ref{sec:results}. 

In order to account for all possible filament axis - los configurations, 
we integrated Eq.~(\ref{eq:cl}) over the inclination angle range 
[$0$,$\pi/2$], assigning each angle a weight $\sin(\theta)$. The 
calculations were carried out in the Rayleigh-Jeans region, for which 
$g(\nu)=-2$, and for the WMAP 7-year cosmological parameter set, 
$\Omega_m=.266,h=.71,n=.963,\sigma_8=.801$, [22]. Due to the very steep 
dependence of the mass function on $\sigma_8$, we explored its 1-$\sigma$ 
uncertainty as well.

Once the temperature anisotropy induced by filaments is 
calculated - our power spectra are shown in Figure 2 and presented in Section 4 - 
it is prudent to assess the bias it introduces in cosmological parameter 
inference, and compare it to the similar cluster-related 
bias; this is discussed in the next Section.

\section{Cosmological Parameter Bias}

The calculated power spectra shown in Figure 2 indicate 
that the contribution of filaments to the power spectrum up to $l\approx 1000$ is 
comparable to that of galaxy clusters. It has been shown in [19] that even after 
removing clusters detected by the PLANCK SZ survey residual contribution to 
the power spectrum is sufficiently large to significantly bias the inferred 
cosmological parameters. Simulated cluster detection and removal that we assume 
for this work (just for the sake of comparison with the filament-induced bias) 
is further discussed in the next section.

One could in principle further reduce this bias by removing the contribution 
of many other clusters not detected by CMB experiments and are detected by 
other means. For example, the Dark Energy Survey (DES) is expected to 
detect $\sim 170,000$ galaxy clusters with a mass limit 
$M_{min}\approx 5\times 10^{13}M_{\odot}$ out to redshift $z\sim 1.5$. Assuming 
that we trust our understanding of the formation of such low mass clusters and 
that our model of virialized clusters is indeed valid, it could be possible 
to significantly suppress the contribution of galaxy 
clusters to the biasing power spectrum to negligible levels. 
This cannot be done with filaments; individual filaments are hard to 
detect (also) in the microwave band due to their only mildly nonlinear density 
contrast. Moreover, unlike clusters these structures contribute at lower 
multipoles, where the CMB is the dominant signal, further challenging their 
detection and removal from CMB maps. 

Even detection and removal by other means (such as measurement of the extremely
low X-ray surface brightness) would be very challenging and obviously uncertain
to the level that their removal might induce unwanted systematics in the data. 
Yet, filaments contribute to the statistical signal, i.e. power spectrum, at a 
level which might bias cosmological parameter estimation as we show below.

Quantitatively, the statistical uncertainty on a given parameter $\lambda_{i}$ 
is retrieved from the Fisher information matrix  
\begin{eqnarray}
F_{ij}=\sum_{l}f_{sky}\left(\frac{2l+1}{2}\right) {\rm Trace}[{\bf C}^{-1}
(\partial {\bf C}_{l}/\partial\lambda_{i}){\bf C}^{-1}(\partial {\bf C}_{l}
/\partial\lambda_{j})]
\label{eq:fisher}
\end{eqnarray}
where for each $l$ the symmetric matrix ${\bf C}$ is
\begin{eqnarray}
{\bf C}_{l} &\equiv& \left(\begin{array}{c c}
C_{l}^{TT} & C_{l}^{TE}\\
C_{l}^{TE} & C_{l}^{EE}\\
\end{array}\right).
\label{eq:matrix}
\end{eqnarray}
The statistical uncertainty on the parameter
$\lambda_{i}$ is $\sigma_{\lambda_{i}}=\sqrt{(({\bf F})^{-1})_{ii}}$. 
Calculating the bias $\delta_{\lambda_{i}}$ (see, e.g., [19])
\begin{eqnarray}
\frac{\delta_{\lambda_{i}}}{\sigma_{\lambda_{i}}}=
-\sum_{l}f_{sky}\sigma_{\lambda_{i}}\left(\frac{2l+1}{2}\right) 
{\rm Trace}[{\bf C}^{-1}(\partial {\bf C}_{l}/\partial\lambda_{i}){\bf C}^{-1}
\Delta {\bf C}_{l}]
\label{eq:bias}
\end{eqnarray}
where here the relative bias $\delta_{\lambda_{i}}/\sigma_{\lambda_{i}}$ 
is with respect to the nominal uncertainty in the parameter $\lambda_{i}$. 
Note that only the (1,1) element of the matrix $\Delta {\bf C}_{l}$ 
is non-vanishing; all other elements vanish as no (even statistically) 
detectable polarization is expected from filaments.

Before we proceed to the results it is important to describe how we actually 
calculated the cluster and filament contribution to $\Delta {\bf C}_{l}$. 
When maps at different frequency bands below and above the SZ crossover 
frequency ($\sim 218$ GHz) are co-added, there is a partial cancellation
of the (dominant, thermal) SZ component. Maps obtained from the various 
frequency channels are weighted by their inverse (measured) power and 
linearly combined into a single power spectrum [20,21]. The weights are
\begin{eqnarray}
(w_{l})_{i}=\frac{\sum_{j}(C_{l}^{-1})_{ij}e_{j}}
{\sum_{i,j}e_{i}(C_{l}^{-1})_{ij}e_{j}}
\label{eq:weight}
\end{eqnarray}
where the indices $i$ and $j$ run over all frequencies, the matrix $C^{-1}$ 
is the inverse total power spectrum, and all components of the vector 
$e$ are identically $1$. The residual foreground power spectrum used in 
Eq.~(\ref{eq:bias}) is then 
$\Delta{\bf C}_{l}=w_{l}^{T}\cdot({\bf C^{-1}})_{l}\cdot w_{l}$.  

\section{Results}
\label{sec:results}
\begin{figure}[t]
\centering
\epsfig{file=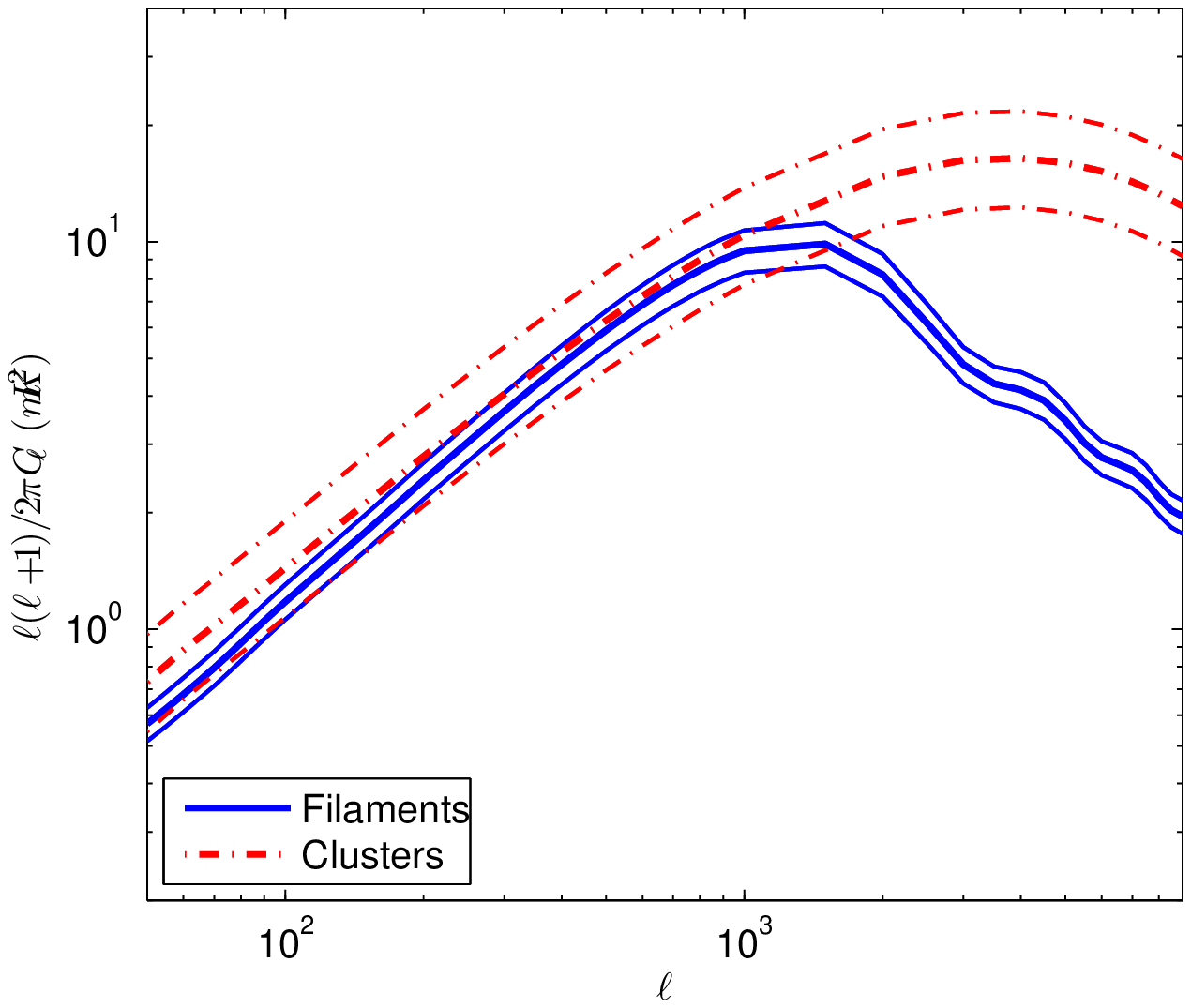, height=7.8cm, width=8cm,clip=}
\hspace{-10mm}
\epsfig{file=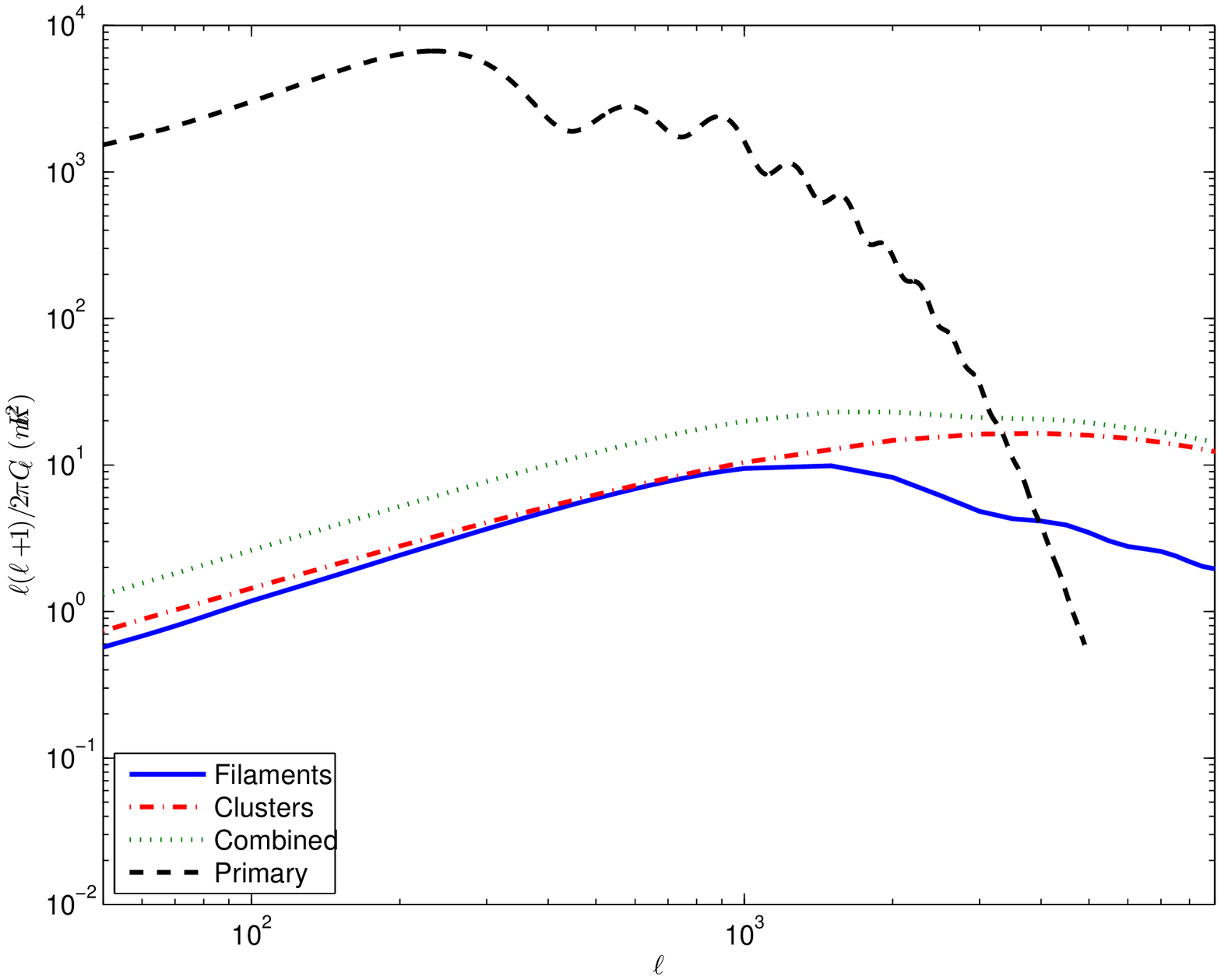, height=7.8cm, width=8cm,clip=}
\caption{Left panel: SZ power spectra due to filaments (continuous blue), 
and clusters (dashed-dotted red), at the RJ limit. The envelopes mark 
the 1-$\sigma$ uncertainty interval $\sigma_8=0.801\pm .03$, as inferred 
from the WMAP 7-year results. Right panel: cluster- and filament-induced 
SZ power plotted together with the primary anisotropy in units of $\mu K^2$. 
The combined SZ power due to clusters and filaments is shown by the dotted 
green curve.}

\label{fig:filsz}
\end{figure}

Results of the calculation are illustrated in the 
left-hand panel of Fig.~\ref{fig:filsz}, where filament-induced SZ power 
levels (marked in blue) are compared with those generated by SZ clusters 
(red). As expected, filament-induced SZ power peaks at lower multipoles 
($\ell\sim 1500$) as compared with that due to clusters ($\ell\sim 4000$), 
owing to the appreciably larger scales involved. With our choice of 
cosmological parameters and IF gas properties, the filaments-induced 
power is comparable to that of clusters for $\ell\lesssim 1000$. The 
oscillating pattern seen at multipoles $\ell\gtrsim 3000$ results from 
the sharp filament edge in our simplified cylindrical model. The combined 
cluster and filament power plotted in the right-hand panel of the figure 
suggests that the filament-induced SZ signal may bias cosmological parameter 
inference from the primary anisotropy (at $\lesssim 1000$) at a similar 
level to that due to clusters.

The procedure for estimation of the statistical uncertainty and bias 
described in Sec. 3 was applied to our 9-parameter cosmological 
model, $A_s$, $\Omega_{\nu}h^{2}$, $\Omega_{m}h^{2}$, 
$\Omega_{b}h^{2}$, $n_{s}$, $h$, $w$, $\tau_{re}$, and $Y_{p}$. In our 
calculations, Eqs.~(\ref{eq:fisher})-(\ref{eq:weight}), we assumed 
a maximum multipole of $l_{max}=3000$. The relevant experiment specifications 
are listed in Table \ref{tab:specs}. Values of the dimensionless bias 
$\delta_{\lambda}/\sigma_{\lambda}$ are listed in Tables \ref{tab:Planck} 
\& \ref{tab:CVL} for Planck and a cosmic-variance-limited experiment (CVL), 
respectively. For each experiment we calculate the bias from filaments 
and the residual cluster signal, assuming that clusters detected at 
$5\sigma$ significance or higher are subtracted out. To estimate the 
cluster detection significance we consider clusters on the mass-redshift 
grid M-z. The gas temperature of a given cluster depends on both its 
redshift and mass assuming it is virialized. In addition, its gas 
density depends on redshift, and therefore the SZ signal from clusters 
depend on their masses and redshift regardless of their abundance, which 
clearly is M-z-dependent. Also, clusters of a given size but at different 
redshifts subtend different angular scales on the sky, a fact relevant 
for their detection because of the telescope limited angular resolution. 

To assess the signal-to-noise of clusters detection we considered noise 
contributions from both the primary CMB, detector noise, and point sources. 
We assume that only data from the 100, 143 and 353 GHz bands is used for 
cluster detection since lower or higher frequency bands are contaminated 
by astrophysical foregrounds (More details are given in, e.g. [23]). 
After removing those `cells' in M-z space where we expect galaxy clusters to 
be detected at 5$\sigma$ or higher, we integrate the M-z-deopendent cluster 
y-parameter over the dark matter halo population, in a  procedure described 
in Eq.(2.12). We assume the Tinker mass function [24]. Finally, to obtain 
the bias on inferred cosmological parameters in the presence of cluster 
residuals (i.e. those clusters that have not been detected and masked 
out), this $\Delta C_{l}$ is used in Eq.(3.3).  

In contrast to galaxy clusters, detection and removal techniques cannot be 
applied to filaments since these irregular low signal structures are not 
likely to be individually detected, almost certainly not at a sufficiently 
high level of significance allowing their removal without biasing the CMB sky. 

It should be mentioned that our results for the expected bias induced by 
{\it galaxy clusters} (Table 2) are systematically lower than those reported 
in [19]. A meaningful comparison between the two sets of results cannot 
be made due to insufficient details on their analysis. A likely source 
for most of the difference is the overall normalization of the SZ power; 
while we normalized to the SPT ($C_{l=3000}^{SZ}=3.65\mu K^{2}$ at 
150 GHz, [25]), their normalization (from Fig.1 in [19]) is considerably 
higher. Also, different procedures of cluster detection and removal 
employed in the two works would almost certainly result in different 
low- and moderate-$l$ power spectrum behavior, this without considering 
the overall different $l$-dependence expected due to the different 
forms of the gas density profile and mass function used in the two works.

It is also interesting to note that in the case of Planck the filaments 
induce bias comparable to the bias induced by galaxy clusters (Table 2), 
and in the case of the CVL experiment the residual cluster-induced bias 
is a factor $\sim 2-3$ larger (Table 3). This implies that the 
contribution of filaments to the bias must be accounted for, especially 
since many more clusters will be detected and removed using non-CMB 
observations, further boosting the relative contribution of 
filaments-induced anisotropy to the bias.

\section{Discussion}

Our calculations of the filament-induced SZ power spectrum are based 
on several simplifying assumptions which merit particular attention, 
given the limited quantitative knowledge of filament structure and 
physical properties.

\noindent
The IF gas temperature and density were taken to be uniform, 
ignoring more likely distributions observed in simulations, with 
higher temperatures and densities lying along the filament axis 
and towards nodes, where clusters are thought to reside. However, 
modeling the distribution of IF gas necessitates a more detailed
study of hydrodynamical simulations and a statistical analysis of 
their results. Our semi-analytic treatment can be readily modified 
once a more detailed description of IF gas properties is obtained 
from hydrodynamical simulations. We do not expect a major revision 
of the results presented here. Note that modification of the IF gas 
temperature is trivially reflected in the SZ power through its direct 
dependence on its squared value. Basically, this holds also for the 
IF electron number density, with the additional effect also on 
the inferred filament length, as apparent from Eq.~(\ref{eq:filh}), 
and consequently also on the shape of the power spectrum, which would 
peak at higher (lower) multipoles with enhanced (reduced) densities. 

\noindent
Likewise, filament morphology is clearly more complex than plain 
cylindrical, and is unlikely to be characterized by a single cross 
section, taken here as the peak distribution value reported in [9]. 
Filament lengths, too, are unlikely to be determined solely by their 
masses; a more plausible calculation would assign lengths drawn from 
a suitable PDF. 

\noindent
Our power spectrum calculations involve the flat sky approximation, 
which formally fails at angular scales associated with long 
filaments. In fact, filaments with mass $10^{15}M_{\odot}\cdot h^{-1}$ 
would be $\sim 140\, Mpc\cdot h^{-1}$ long, and would subtend an angle 
$\sim 22^{\circ}$ at redshift $z=0.1$. We nonetheless believe that our 
results are still viable, since the contribution of massive filaments 
to the SZ power spectrum is quite low due to their extreme rarity, as 
predicted by the mass function, whereas they would induce the maximal 
SZ signal at the least probable low inclination angles with the los, 
for which the flat sky approximation holds. At the largest (and most 
probable) inclination angle of $90^{\circ}$, the flat sky approximation 
may seem insufficiently accurate; however, at such angles the optical 
depth is comparatively small and the SZ signal is relatively weak. 
Moreover, to obtain a quantitative estimate for the departure of the 
flat-sky from the full-sky calculation at the lowest multipoles, we note 
that a first order estimate of the accuracy of the approximation could 
be obtained by realizing that going from flat- to full-sky may be 
approximated by either taking the limit $l\rightarrow l+\frac{1}{2}$ or 
$l\rightarrow\sqrt{l(l+1)}$ [26]. This implies that by employing the 
flat-sky approximation we systematically underestimate the filament 
power spectrum at the lowest multipoles, since the filament power 
spectrum has a positive slope at $l\lesssim 1000$ (Fig.[2]). Now, 
beacuase the bias, Eq.(3.3), is proportional to the filament power 
spectrum this implies that we actually underestimate the level of 
cosmological parameter bias. Employing these transformations in our 
calculations shows that the values reported in Tables 2 \& 3 change 
only at the third or fourth decimal point. This can be explained by 
the fact that the flat sky approximation only makes a noticeable 
difference at very low multipoles. Indeed, we verified that already 
at $l=10$ there is only $\sim 3-4\%$ departure, smaller than 
cosmic-variance at this scale. Compared to other uncertainties in our 
filament model, the impact of this approximation on our bias estimates 
is negligibly small.
\newline
\noindent

\section{Conclusions}

We used a simple model for the intra-filament gas to assess SZ power 
generated in these large scale structures. We have shown that under 
certain simplifying assumptions, SZ power induced by filaments could 
reach a level of $\sim 10 \mu K^{2}$ at multipoles $\ell\sim 1500$ in 
the Rayleigh-Jeans region, with uncertainties directly related to IF 
gas properties and morphology. SZ power induced by filaments at 
relevant scales is shown to constitute a non-negligible contribution to 
the overall SZ signal, with respect to the corresponding power generated 
by SZ clusters. Unlike clusters, whose SZ signal can be detected and 
removed from the CMB sky without introducing appreciable bias, the 
diffuse filament SZ signal cannot be subtracted out. 
Consequently, the SZ power contributed by filaments can potentially 
induce a non-negligible bias on several cosmological parameters as shown 
in Tables \ref{tab:Planck} \& \ref{tab:CVL}. 

The bias levels shown in Tables \ref{tab:Planck} \& \ref{tab:CVL} 
were obtained from using all 3000 multipoles. We note that 
by cutting off at lower $l_{max}$ values the bias decreases. 
For PLANCK we find that already at $l_{c}\approx 1000$ both 
$A_{s}$, $n_{s}$ and $w$ are biased by a level comparable to 
the nominal statistical uncertainty. With the CVL experiment 
this critical multipole number is pushed towards $\approx 1500$.

It is interesting to note that the pervasive web of filamentary structure 
represents the most extended screen that sets a lower limit to the degree 
of Comptonization of the CMB, a limit that cannot be removed when 
attempting to use spectral distortions of the CMB as cosmological probes 
of energy release processes in the early universe. From the filament-induced 
temperature anisotropy power at the lowest 2000-3000 multipoles we estimate 
an rms comptonization parameter $y_{rms}\sim 7\times 10^{-7}$. This value 
scales approximately as $\propto(n_{e}T_{e})^{2}$. Due to modeling 
uncertainties - e.g. filament morphology, filament mass function, gas 
density and temperature - there clearly is an appreciable uncertainty 
in our estimate for $y_{rms}$. However, since the missing baryons are 
likely to be accounted for by warm IF gas, $n_{e}$ is not expected to 
significantly deviate from the fiducial value adopted here because of 
the constraint set by the requirement that the missing baryons provide 
30-40\% of the total baryon budget. Ultimately, these uncertainties 
will be lowered through more realistic high resolution hydrodynamical 
simulations.

\acknowledgments

This work was supported by 
a US-IL Binational Science Foundation grant 2008452, and by the 
James B. Ax Family Foundation. We thank two anonymous referees 
for useful comments. Use of CAMB for calculations of primary CMB 
power spectra is acknowledged.
\newpage

\begin{table}
\begin{tabular}{|c|c|c|c|c|}
\hline
Experiment & $f_{\rm sky}$ & $\nu [GHz]$ & $\theta_b [1']$ & $\Delta_T [\mu K]$\\
\hline
& &  30 & 33 &  4.4\\
& &  44 & 23 &  6.5\\
& &  70 & 14 &  9.8\\
& &  100 & 9.5 &  6.8\\
Planck&0.65& 143 & 7.1 & 6.0\\
& &  217 & 5.0 &  13.1\\
&   & 353 & 5.0 & 40.1\\
& &  545 & 5.0 &  401\\
&   & 856 & 5.0 & 18300\\
\hline
\hline
\end{tabular}
\caption{Sensitivity parameters for Planck. The CVL experiment is noiseless. 
The beam full width half maximum (FWHM) resolution $\theta_{b}$ is related to 
the gaussian beamwidth via $\theta_{b}=\sqrt{8\ln(2)}\theta_{b}$ and the detector 
noise power spectrum is $C_{l}^{T,det}=(\Delta_{T}\theta_{b})^{2}e^{l^{2}\sigma_{b}^{2}}$. 
The figures quoted here assume 1 year of observation [27].}
\label{tab:specs}
\end{table}

\begin{table}
\begin{tabular}{|c|c|c|c|c|c|c|c|c|c|c|}
\hline
Foreground &$A_{s}$ & $\Omega_{\nu}h^{2}$ & $\Omega_{m}h^{2}$ & $\Omega_{b}h^{2}$ 
& $n_{s}$ & $h$ & $w$ & $\tau$ & $Y_{p}$\\
\hline
Filaments &$1.5$&$-0.8$&$-0.3$&$0.0$&$1.8$&$-0.8$&$-1.3$&$-0.6$&$-0.3$\\
\hline
Cluster Residuals & $1.1$&$-0.8$&$-0.3$&$0.1$&$1.4$&$-0.7$&$-1.1$&$-0.4$&$-0.3$\\
\hline
\end{tabular}
\caption{Expected bias of the cosmological model due to an unaccounted-for 
filament or cluster contribution to the CMB power spectrum measured by Planck 
in units of the nominal statistical uncertainty $\sigma_{\lambda}$. 
In the case of filaments the assumed electron number density and gas 
temperature are $n_{0}=4.3\times 10^{-6} cm^{-3}$ and $kT_{e}=1 keV$, 
respectively. The first row shows the bias due to CMB comptonization by 
the warm plasma in filaments. In the second row we present the effect of 
the thermal SZ effect from galaxy clusters. Shown is the reduced bias level 
that would be obtained if the clusters detected by Planck would be removed 
from the CMB sky.} 
\label{tab:Planck}
\end{table}

\begin{table}
\begin{tabular}{|c|c|c|c|c|c|c|c|c|c|c|}
\hline
Foreground &$A_{s}$ & $\Omega_{\nu}h^{2}$ & $\Omega_{m}h^{2}$ & $\Omega_{b}h^{2}$ 
& $n_{s}$ & $h$ & $w$ & $\tau$ & $Y_{p}$\\
\hline
Filaments &$5.6$&$-7.5$&$-1.9$&$0.7$&$6.5$&$-6.2$&$-8.4$&$-2.6$&$-1.6$\\
\hline
Cluster Residuals & $13.4$&$-17.6$&$-4.4$&$1.7$&$13.7$&$-11.7$
&$-15.4$&$-5.2$&$-4.0$\\
\hline
\end{tabular}
\caption{Same as Table 2, but for a CVL experiment} 
\label{tab:CVL}
\end{table}

\end{document}